\documentclass[reprint,superscriptaddress,amsmath,amssymb,aps,prl,longbibliography]{revtex4-1}

\usepackage{amssymb,amsmath}
\usepackage{cmap}
\usepackage{graphicx}
\usepackage{bm}
\usepackage{color}
\usepackage{blindtext}
\usepackage{hyperref}
\usepackage{listings}
\usepackage{booktabs}
\usepackage{multirow}
\usepackage{amsmath}
\usepackage{mathrsfs}
\usepackage{braket}

\usepackage{dsfont}
\usepackage{tabularx}

\renewcommand{\vec}[1]{\boldsymbol{\mathbf{#1}}}

\renewcommand{\Re}{\operatorname{Re}}
\renewcommand{\Im}{\operatorname{Im}}

\begin{document}

\title{Acoustic Radiation Force and Torque on Small Particles \\ as Measures of the Canonical Momentum and Spin Densities}

\newcommand{\affilITMO}{ITMO University, Birzhevaya liniya 14, St.-Petersburg 199034, Russia}
\newcommand{\affilRIKEN}{Theoretical Quantum Physics Laboratory, RIKEN Cluster for Pioneering Research, Wako-shi, Saitama 351-0198, Japan}
\newcommand{\affilANU}{Nonlinear Physics Centre, RSPE, The Australian National University, Canberra, ACT 0200, Australia}

\author{I. D. Toftul}
\affiliation{\affilRIKEN}
\affiliation{\affilITMO}

\author{K. Y. Bliokh}
\affiliation{\affilRIKEN}
\affiliation{\affilANU}

\author{M. I. Petrov}
\affiliation{\affilITMO}

\author{F. Nori}
\affiliation{\affilRIKEN}
\affiliation{Physics Department, University of Michigan, Ann Arbor, Michigan 48109-1040, USA}


\begin{abstract}
We examine acoustic radiation force and torque on a small (subwavelength) absorbing isotropic particle immersed in a monochromatic (but generally inhomogeneous) sound-wave field. We show that by introducing the monopole and dipole polarizabilities of the particle, the problem can be treated in a way similar to the well-studied optical forces and torques on dipole Rayleigh particles. We derive simple analytical expressions for the acoustic force (including both the gradient and scattering forces) and torque. Importantly, these expressions reveal intimate relations to the fundamental field properties introduced recently for acoustic fields: the canonical momentum and spin angular momentum densities. We compare our analytical results with previous calculations and exact numerical simulations. We also consider an important example of a particle in an evanescent acoustic wave, which exhibits the mutually-orthogonal scattering (radiation-pressure) force, gradient force, and torque from the transverse spin of the field. 	
\end{abstract}


\maketitle

{\it Introduction.---}
Optical and acoustic radiation forces and torques are of great importance from both practical and fundamental points of view. On the one hand, these mechanical manifestations of the radiation power underpin optical and acoustic manipulations of small particles \cite{Ashkin2000,Grier2003,Dienerowitz2008,Sukhov2017,Laurell2007,
Ding2012}, atomic cooling \cite{Stenholm1986,Chu1998,Phillips1998}, optomechanics \cite{Aspelmeyer2014}, acoustofluidics \cite{Bruus2012a,Ding2013}, etc. On the other hand, radiation forces and torques reveal the fundamental {\it momentum} and {\it angular-momentum} properties of the optical and sound wave fields \cite{Jones1953,Loudon2012,Brevik1979,Pfeifer2007,He1995,ONeil2002,Garces2003,
Volke2008,Zhang2011,Demore2012,Anhauser2012}. 

Since Kepler's observation of the comet tail and early theoretical works by Euler and Poynting \cite{Jones1953,Loudon2012}, the studies of optical and acoustic momentum and forces were developed in parallel ways. Remarkably, despite numerous works calculating radiation forces and torques acting on various small particles in optics \cite{Askaryan1962,Gordon1973,Ashkin1983,Marston1984,Nieto-Vesperinas2000,Nieto-Vesperinas2010,Ruffner2012} and acoustics \cite{Gorkov1962,Hasegawa1977a,Busse1981,Doinikov1994,Settnes2012,Silva2014,Karlsen2015,
Zhang2014,Silva2013,Zhang2018a,Fan2019}, the explicit proportionality of the force and torque to the local wave momentum and spin angular momentum densities was properly established {\it in optics} only recently \cite{Berry2009,Bliokh2013,Genet2013,Bliokh2014,Bliokh2014_II,Bekshaev2015,
Bliokh2015,Nieto-Vesperinas2015,Leader2016,Antognozzi2016}. The reason for this is that, in generic inhomogeneous wave fields, the force and torque on an isotropic small absorbing particle are proprtional to the {\it canonical} momentum and spin densities rather than the Poynting (kinetic) momentum and angular momentum commonly used for many decades \cite{Bliokh2014,Bliokh2014_II,Bekshaev2015,Bliokh2015,Leader2016,Bliokh2013_II,
Leader2014,Aiello2015}. 

In {\it acoustics}, such explicit connection between the radiation force/torque and momentum/spin in generic inhomogeneous fields has not been described so far. Moreover, the concepts of the {\it canonical momentum} and {\it spin angular momentum} densities in sound wave fields have been introduced only in very recent works \cite{Long2018,Shi2019,Bliokh2019a,Bliokh2019}. 

In this work, we provide a simple yet accurate theory of acoustic forces and torques on small (subwavelength) absorbing isotropic particles in generic monochromatic acoustic fields. By employing methods well-established in optics and involving the monopole and dipole {\it polarizabilities} of the particle (determined by the leading terms in the Mie scattering problem), we derive simple analytical expressions for the acoustic forces and torque. Most importantly, these expressions indeed expose the intimate relation to the canonical momentum and spin densities in the acoustic field. We show that our results agree with specific previous calculations and exact numerical simulations. We illustrate our general theory with an explicit example of the forces and torque on a small particle in an evanescent acoustic wave.

{\it Properties of acoustic fields.---}
We will deal with monochromatic but arbitrarily inhomogeneous acoustic fields of frequency $\omega$ in a homogeneous dense medium (fluid or gas). The complex pressure and velocity fields, $p({\bf r})$ and ${\bf v({\bf r})}$, obey the wave equations \cite{landau1982fluid}
\begin{equation}
\label{Eq.1}
i\omega\beta\, p =  \nabla  \cdot {\bf{v}}~,\quad
i\omega \rho\, {\bf{v}} = \nabla p~,
\end{equation}
where the medium is characterized by the compressibility $\beta$, the mass density $\rho$, and the speed of sound $c=1/\sqrt{\rho\beta}$. 

We will characterize the dynamical properties of the acoustic wave field by its energy,  {\it canonical} momentum, and spin angular momentum densities. The energy density reads \cite{landau1982fluid}:
\begin{equation}
\label{Eq_energy}
W = \frac{1}{4}\left( {\beta {|p|^2} + \rho {|{\bf v}|^2}}\right) \equiv
W^{(p)} + W^{({\bf v})}.
\end{equation}
The canonical momentum and spin densities of acoustic fields were introduced very recently \cite{Shi2019,Bliokh2019a,Bliokh2019}:
\begin{eqnarray}
{\bf P} &=& \frac{1}{4\omega} \Im\! \left[ \beta\, p^*\nabla p + \rho\, {\bf v}^*\!\cdot (\nabla) {\bf v} \right]  
\equiv {\bf P}^{(p)} + {\bf P}^{({\bf v})}, 
\label{Eq_momentum} \\	
{\bf S} &=& \frac{\rho}{2\omega} \Im\! \left( \vec{v}^*\! \times \vec{v} \right),
\label{Eq_spin}
\end{eqnarray}
where $\left[ {\bf v}^*\!\cdot (\nabla) {\bf v} \right]_i \equiv \Sigma_j v_j^{*} \nabla_i v_j$ \cite{Berry2009}.

The energy (\ref{Eq_energy}) and momentum (\ref{Eq_momentum}) densities are represented as symmetric sums of the pressure- and velocity-related contributions, indicated by the corresponding superscripts. This is similar to the symmetric electric- and magnetic-field contributions in electromagnetism \cite{Berry2009,Bliokh2013_II,Bliokh2014,Bliokh2014_II,Bekshaev2015,Bliokh2015,
Cameron2012,Aiello2015}. In contrast, the spin density (\ref{Eq_spin}) has only the velocity contribution because the {\it scalar} pressure field cannot generate any local vector rotation.  

Note that the canonical momentum determines the orbital angular momentum density ${\bf L} = {\bf r} \times {\bf P}$ \cite{Bliokh2014,Bliokh2015,Bliokh2013_II,Leader2014,Bliokh2019}, and that the more familiar {\it kinetic} momentum density (the acoustic analogue of the Poynting momentum) is given by ${\bm \Pi} = {\bf P} + \dfrac{1}{4} \nabla \times {\bf S} = \dfrac{1}{2c^2}\Re(p^* {\bf v})$ \cite{landau1982fluid,Bliokh2019}. The equivalence of the canonical and kinetic momentum and angular momentum quantities appears for their integral values for localized acoustic fields: $\langle {\bf P} \rangle = \langle {\bm \Pi} \rangle$ and $\langle {\bf S} \rangle + \langle {\bf L} \rangle = \langle {\bf r} \times {\bm \Pi} \rangle$ \cite{Bliokh2014,Bliokh2015,Bliokh2013_II,Leader2014,Bliokh2019}, where the angular brackets stand for spatial integration.  
However, here we are interested in {\it local} rather than integral field properties, which are very different in the canonical and kinetic pictures; below we show that it is the {\it canonical} quantities (\ref{Eq_momentum}) and (\ref{Eq_spin}) which correspond to the force and torque on small particles. 

{\it Interaction with a small particle.---}
The most straightforward way to detect the momentum and angular momentum of a wave field is to measure the force and torque it produces on a probe particle \cite{Jones1953,Loudon2012,Brevik1979,Pfeifer2007,He1995,ONeil2002,Garces2003,
Volke2008,Zhang2011,Demore2012,Bliokh2013,Genet2013,Bliokh2014,Bliokh2014_II,
Bliokh2015,Nieto-Vesperinas2015,Leader2016,Antognozzi2016,Liu2018,Shi2019}. Therefore, we consider the interaction of a monochromatic acoustic wave with a small (subwavelength) spherical isotropic particle of the radius $a$, density $\rho_1$ and compressibility $\beta_1$, with its center at ${\bf r}={\bf r}_0$. We allow the particle to be {\it absorbing}, i.e., the parameters $\{\rho_1,\beta_1 \}$ are generally {\it complex}. 

The wave-particle interaction is directly related to the wave scattering problem. For small isotropic particles, the scattered field is conveniently represented by a multipole expansion \cite{bohren2008absorption,williams1999fourier,blackstock2001fundamentals}, where the small parameter is $ka\ll 1$ ($k=\omega/c$ is the wave number). For electromagnetic waves, the leading term is the {\it dipole} one \cite{Ashkin2000,Grier2003,Dienerowitz2008,Sukhov2017,Askaryan1962,Gordon1973,
Ashkin1983}, because the monopole cannot radiate transversal waves. In contrast, for longitudinal acoustic waves, the leading terms are the {\it monopole and dipole} ones, and these generally have the same order in $ka$ \cite{Gorkov1962,Bruus2012a}. Therefore, a small particle in an acoustic wave field can be approximated by a monopole and dipole, which are induced by the incident field and are interacting with this field (so the interaction is quadratic with respect to the field). 

\begin{figure}
	\centering
	\includegraphics[width=0.85\linewidth]{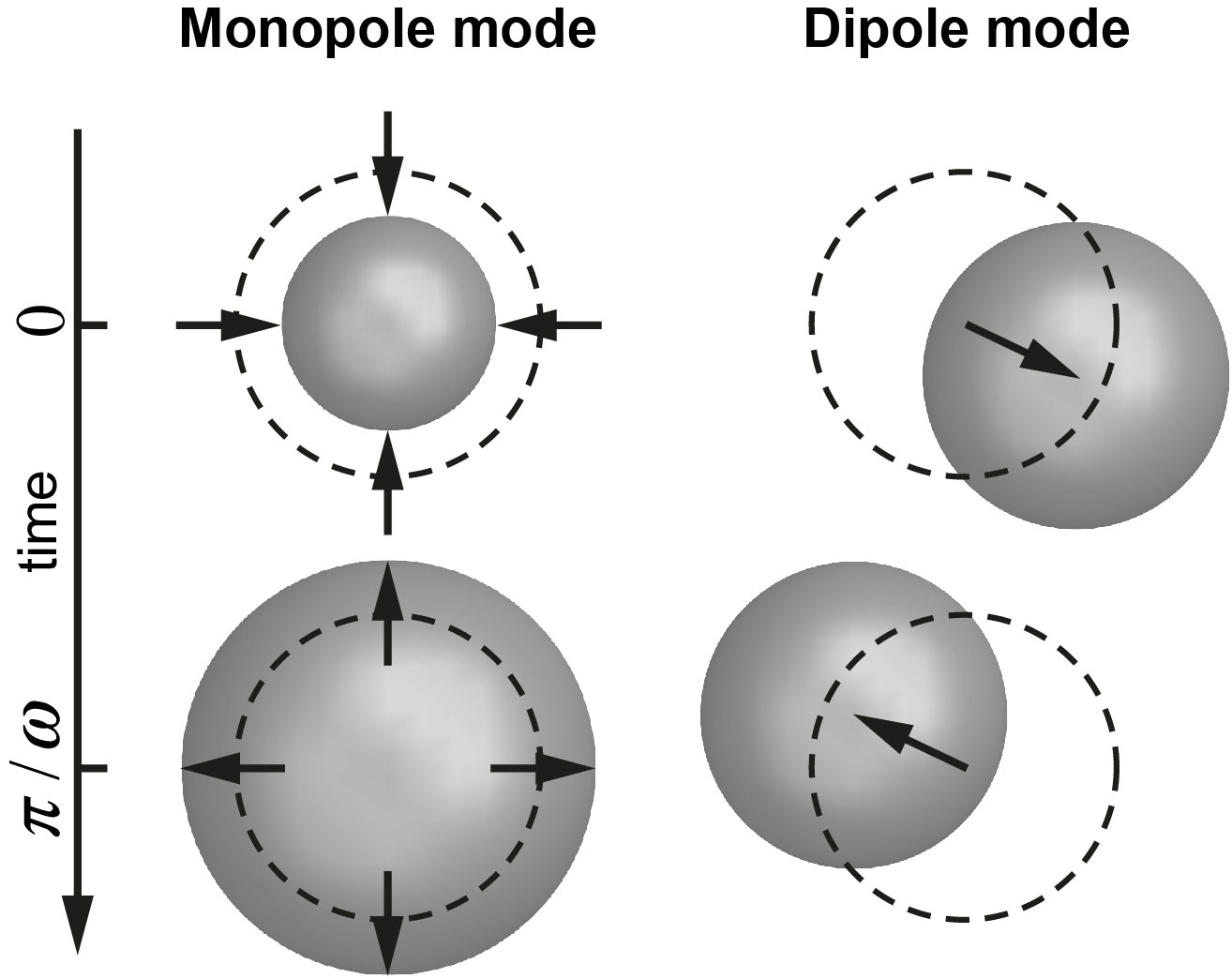}
	\caption{The monopole and dipole oscillatory modes of a spherical particle. These modes are associated with an isotropic compression/expansion and a linear oscillatory motion of the particle, which are induced by the oscillating scalar pressure $p$ and vector velocity ${\bf v}$ fields, respectively.}
	\label{Fig1}
\end{figure}

The oscillating monopole and dipole modes of the particle are schematically shown in Fig.~\ref{Fig1}. The monopole mode is associated with the isotropic compression/expansion of the sphere, while the dipole mode represents oscillations of the particle position along certain direction. It is easy to see that these modes can be excited by the oscillating pressure $p$ and velocity ${\bf v}$ fields, respectively. Therefore, the induced monopole and dipole moments of the particle can be written as:
\begin{equation}
Q = -i\,\omega\beta\, \alpha_m\, p({\bf r}_0)~, \qquad \vec{D} = \alpha_d\, \vec{v}({\bf r}_0)~,
\label{Eq_moments}
\end{equation}
where, following optical terminology, $\alpha_m$ and $\alpha_d$ are the monopole and dipole {\it polarizabilities} of the particle, and the prefactor $-i\,\omega\beta$ in the monopole term is introduced for the convenience in what follows and equal dimensionality of the polarizabilities. Comparing the leading terms of the multipole expansion of the acoustic Mie scattering problem with the standard expressions for the acoustic monopole and dipole radiation \cite{williams1999fourier,blackstock2001fundamentals}, we find the expressions for the polarizabilities (see Supplemental Material \cite{SM}):
\nocite{Yosioka1955, Doyle1989, Jylha2006, Moroz2009, Evlyukhin2010, Evlyukhin2012, LeRu2013, Westervelt1951, Livett1981, Maidanik1958, Zhang2011a, Maidanik1957, Mitri2007, Gouesbet2011, Silva2011, Silva2012}
\begin{eqnarray}
\alpha_m = -\frac{4\pi i}{k^3} a_0 \simeq \frac{4\pi}{3}a^3 \left( \bar{\beta} - 1 \right),
\nonumber\\
\alpha_d = -\frac{4\pi i}{k^3} 3 a_1 \simeq \frac{4\pi}{3} a^3\, \frac{3(\bar{\rho} - 1)}{2\bar{\rho} +1}\, ,
\label{Eq_polarizabilities}
\end{eqnarray}
Here, $\bar{\rho} = \rho_1 / \rho$ and $\bar{\beta} = \beta_1 / \beta$ are the relative density and compressibility of the particle, $a_0$ and $a_1$ are the first two Mie scattering coefficients, and we approximated these coefficients by the leading $(ka)^3$ term in $ka \ll 1$ (see Supplemental Material \cite{SM}). Naturally, the monopole and dipole polarizabilities are related to the differences in the compressibilities and mass densities between the particle and the surrounding medium, respectively. These differences induce relative compression and shift of the particle as shown in Fig.~\ref{Fig1}. 

{\it Absorption rate, force, and torque.---}
The interaction of the induced monopole and dipole moments of the particle with the acoustic field can be described via the minimal-coupling model between the moments (\ref{Eq_moments}) $(Q,{\bf D})$ and the fields $(p,{\bf v})$. Introducing the proper dimensional coefficients, the complex interaction energy takes the form $W^{\rm int}\! = \dfrac{1}{2}\left(\dfrac{i}{\omega}\, Q^* p - \rho\, {\bf D}^* \cdot {\bf v} \right)$. Notably, this energy is precisely equivalent to the energy of the electric dipole ${\bf D}$ and charge $Q$ in the electric field ${\bf E} = \rho\, {\bf v}$ and the corresponding electric potential $\Phi = i\,\omega^{-1} p$ (${\bf E} = - \nabla \Phi$).

The interaction can be characterized by the rates of the {\it energy}, {\it momentum}, and {\it angular momentum} transfer between the field and the particle, which are quantified by the {\it absorption rate}, {\it radiation force}, and {\it radiation torque}, respectively \cite{Bliokh2014_II}. First, the absorption rate is determined by the imaginary part of the interaction energy:
\begin{equation}
	\mathscr{A} = \omega \Im (W^{\rm int}) = 2\,\omega\! \left[ \Im (\alpha_m)\, W^{(p)}\! + \Im (\alpha_d)\, W^{({\vec v})} \right]\!.
	\label{Eq_absorption}
\end{equation}
It is naturally proportional to the imaginary parts of the particle polarizabilities (\ref{Eq_polarizabilities}) (and, hence, of the parameters $\bar{\rho}$ and $\bar{\beta}$) and to the corresponding pressure- and velocity-related energy densities (\ref{Eq_energy}) of the field. 

Second, the radiation force is associated with the gradient of the real part of the interaction energy and can be written as \cite{Dienerowitz2008,Sukhov2017,Ashkin1983,Nieto-Vesperinas2000,Nieto-Vesperinas2010,Bliokh2014_II}:
\begin{eqnarray}
\hspace*{-0.4cm}
{\bf F} =\! - \frac{1}{2} \Re\! \left[ \frac{i}{\omega}\, Q^* \nabla p - \rho\, {\bf D}^*\! \cdot (\nabla) {\bf v} \right]\! = {\bf F}^{\rm grad}\! + {\bf F}^{\rm scat}.
\label{Eq_force}
\end{eqnarray}
Here the gradient and scattering parts are related to the real and imaginary parts of the particle polarizabilities:
\begin{eqnarray}
	{\bf F}^{\rm grad} = \Re (\alpha_m) \nabla W^{(p)} + \Re (\alpha_d) \nabla W^{({\vec v})} ,	
\label{Eq_gradient} \\
{\bf F}^{\rm scat} = 2\,\omega\! \left[ \Im (\alpha_m)\, {\vec P}^{(p)} + \Im (\alpha_d)\, {\vec P}^{({\vec v})} \right] .	
\label{Eq_scattering}
\end{eqnarray}
These laconic expressions reveal the direct relation between the scattering force (which is associated with the absorption of phonons by the particle) and {\it canonical momentum} density (\ref{Eq_momentum}) of the acoustic field. Importantly, substituting the polarizabilities (\ref{Eq_polarizabilities}) into Eqs.~(\ref{Eq_gradient}) and (\ref{Eq_scattering}), one can check that the gradient force exactly coincides with the force found in earlier calculations for lossless particles \cite{Bruus2012a,Gorkov1962,Silva2014,Fan2019} (${\bf F}^{\rm scat} = 0$ in this approximation), while the scattering-force part is equivalent to that found in recent works \cite{Settnes2012,Karlsen2015} considering viscous fluids.
Remarkably, Eqs.~(\ref{Eq_force})--(\ref{Eq_gradient}) are entirely similar to the expressions for optical radiation forces on small Rayleigh particles or atoms \cite{Dienerowitz2008,Sukhov2017,Askaryan1962,Gordon1973,Ashkin1983,Nieto-Vesperinas2010,Ruffner2012,Berry2009,Bliokh2013,Genet2013,Bliokh2014,Bekshaev2015,
Bliokh2015,Bliokh2014_II,Nieto-Vesperinas2000}. In this manner, the electric- and magnetic-dipole terms in optical equations \cite{Bliokh2014,Bliokh2015,Bliokh2014_II,Bekshaev2015,Nieto-Vesperinas2000} (related to the electric and magnetic fields ${\bf E}$ and ${\bf H}$) correspond to the monopole and dipole terms in the acoustic equations (related to the pressure and velocity fields $p$ and ${\bf v}$).

Using the above correspondence between the optical and acoustic interactions, we readily find the acoustic torque on a small particle. The isotropic monopole moment cannot induce any torque, and the torque originates solely from the dipole moment ${\bf D}$ of the particle. In analogy with an electric dipole in an electric field \cite{Genet2013,Bliokh2014,Bliokh2015,Bliokh2014_II,Nieto-Vesperinas2015}, we obtain:
\begin{eqnarray}
	{\bf T} = \frac{1}{2} \Re \left[ \rho\, {\bf D}^*\! \times {\bf v} \right] = 
\omega \Im (\alpha_d)\, {\bf S}\,.
	\label{Eq_torque}
\end{eqnarray}
The very simple Eq.~(\ref{Eq_torque}) reveals the direct connection between the {\it spin angular momentum} density (\ref{Eq_spin}) of the acoustic field and the radiation torque on a small absorptive particle. To the best of our knowkedge, this equation has not been derived before. This general connection (entirely similar to the optical case) is very important, because it was implied without rigorous grounds in the very recent experiment measuring acoustic spin \cite{Shi2019}. Furthermore, this connection can be seen by comparing very recent numerical simulations of the acoustic torque and analytical calculations of the spin density in the particular case of acoustic Bessel beams \cite{Zhang2018a,Bliokh2019}. Having the simple expression (\ref{Eq_torque}), acoustic torques on subwavelength isotropic particles can be readily found analytically in an {\it arbitrary} acoustic field.

\begin{figure}
\centering
\includegraphics[width=0.9\linewidth]{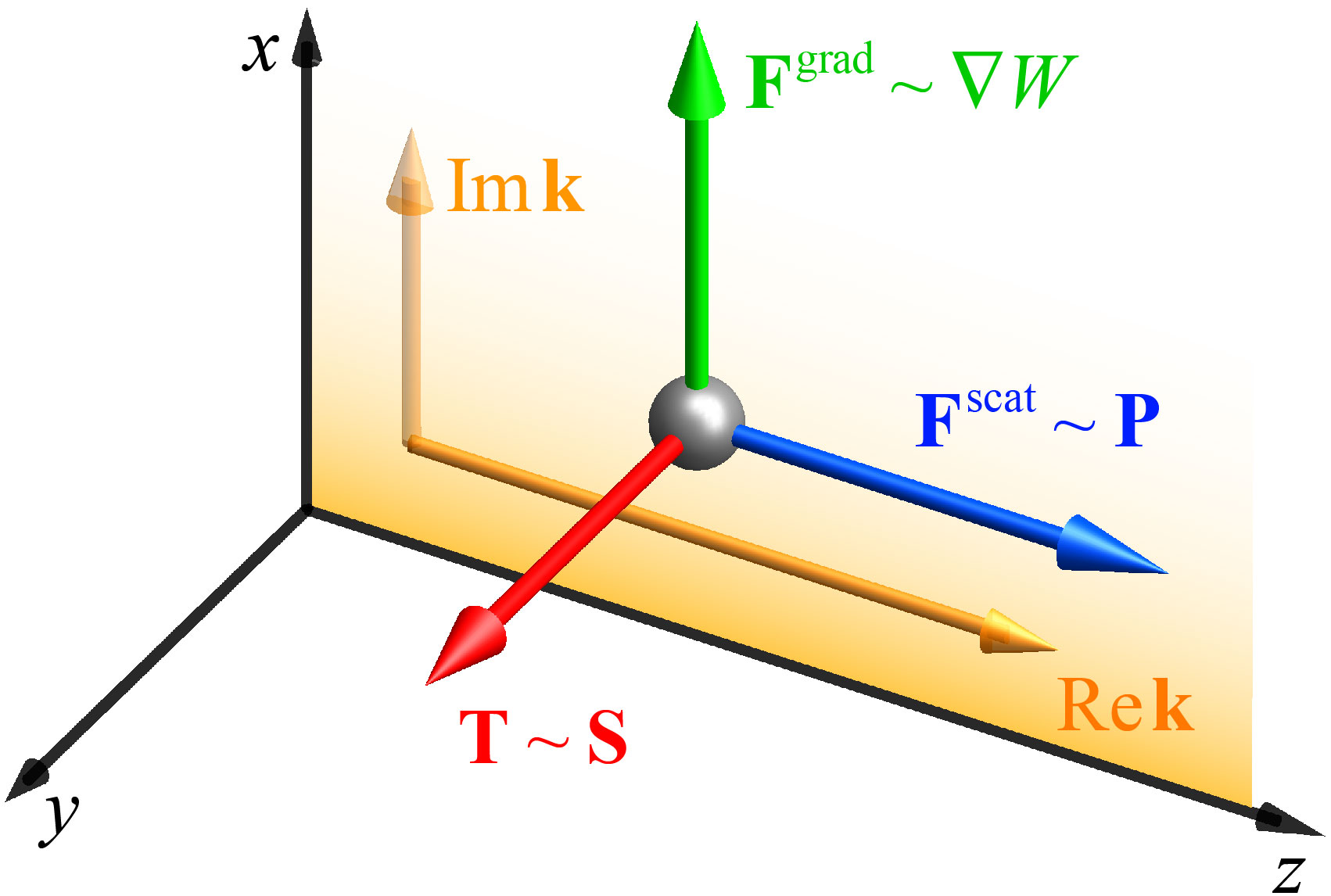}
\caption{A small spherical particle in the acoustic evanescent field (\ref{Eq_evanescent}), which can be treated as a plane wave with the complex wave vector ${\bf k} = k_z \bar{\bf z} + i \kappa\, \bar{\bf x}$. The gradient and scattering (radiation pressure) forces (\ref{Eq_gradient}) and (\ref{Eq_scattering}) are produced by the energy density gradient and canonical momentum (the real part of the wave vector), respectively. The torque (\ref{Eq_torque}) is produced by the transverse spin of the evanescent field \cite{Bliokh2014,Bliokh2015,Aiello2015,Shi2019,Bliokh2019a}.}
\label{Fig2}
\end{figure}

Equations (\ref{Eq_absorption})--(\ref{Eq_torque}) are the main results of our work. Even though some of these are equivalent to the previously-known expressions (such as gradient force on lossless particles), here the acoustic absorption, forces, and torque are for the first time presented in a unified and physically clear form. All these quantities are determined by the fundamental energy, momentum, and angular-momentum properties (\ref{Eq_energy})--(\ref{Eq_spin}) of the field, as well as by the monopole and dipole particle polarizabilities (\ref{Eq_moments}) and (\ref{Eq_polarizabilities}). Note that all the quantities (\ref{Eq_polarizabilities})--(\ref{Eq_torque}) behave as $\propto(ka)^3$, i.e., proportionally to the {\it volume} of the particle. This makes perfect physical sense and allows one to discriminate between various calculations of radiation forces and torques (see, e.g., torques in \cite{Busse1981,Silva2013} with dependences $\propto (ka)^2$ and $\propto (ka)^5$, respectively). For larger or lossless ($\Im(\alpha_{m,d})=0$) particles, one has to involve higher-order terms in $ka$ (see below).

{\it Example: Forces and torques in an evanescent acoustic field.---}
To illustrate the above general theory, we consider a single evanescent acoustic wave with the pressure and velocity fields given by \cite{Shi2019,Bliokh2019a}:
\begin{equation}
p = A\, e^{ik_z z - \kappa x}, \quad \vec{v} = \frac{A}{\omega \rho} \begin{pmatrix}
i\kappa \\ 0 \\ k_z
\end{pmatrix} e^{ik_z z - \kappa x}.
\label{Eq_evanescent}
\end{equation}
Here, $A$ is a constant amplitude, $k_z$ is the longitudinal propagation constant, and  $\kappa$ is the vertical decay constant. This example is very simple yet generic. On the one hand, the evanescent wave can be treated as a {\it plane wave} with the {\it complex} wave vector ${\bf k} = k_z \bar{\bf z} + i \kappa\, \bar{\bf x}$ (the overbar denotes the unit vectors of the corresponding axes) \cite{Bliokh2014,Bliokh2015}, Fig.~\ref{Fig2}, which allows one to use the exactly solvable Mie scattering problem for numerical calculations of forces and torques \cite{bekshaev2013mie}. On the other hand, the evanescent wave is {\it inhomogeneous}, and it carries the intensity gradient $\nabla W$, canonical momentum ${\bf P}$, and spin ${\bf S}$, which exert the gradient force (\ref{Eq_gradient}), scattering forces (\ref{Eq_scattering}), and the radiation torque (\ref{Eq_torque}) in the three mutually-orthogonal directions \cite{Bliokh2014,Bliokh2015,Aiello2015,Shi2019,Bliokh2019a}, see Fig.~\ref{Fig2}.

\begin{figure}
\centering
\includegraphics[width=\linewidth]{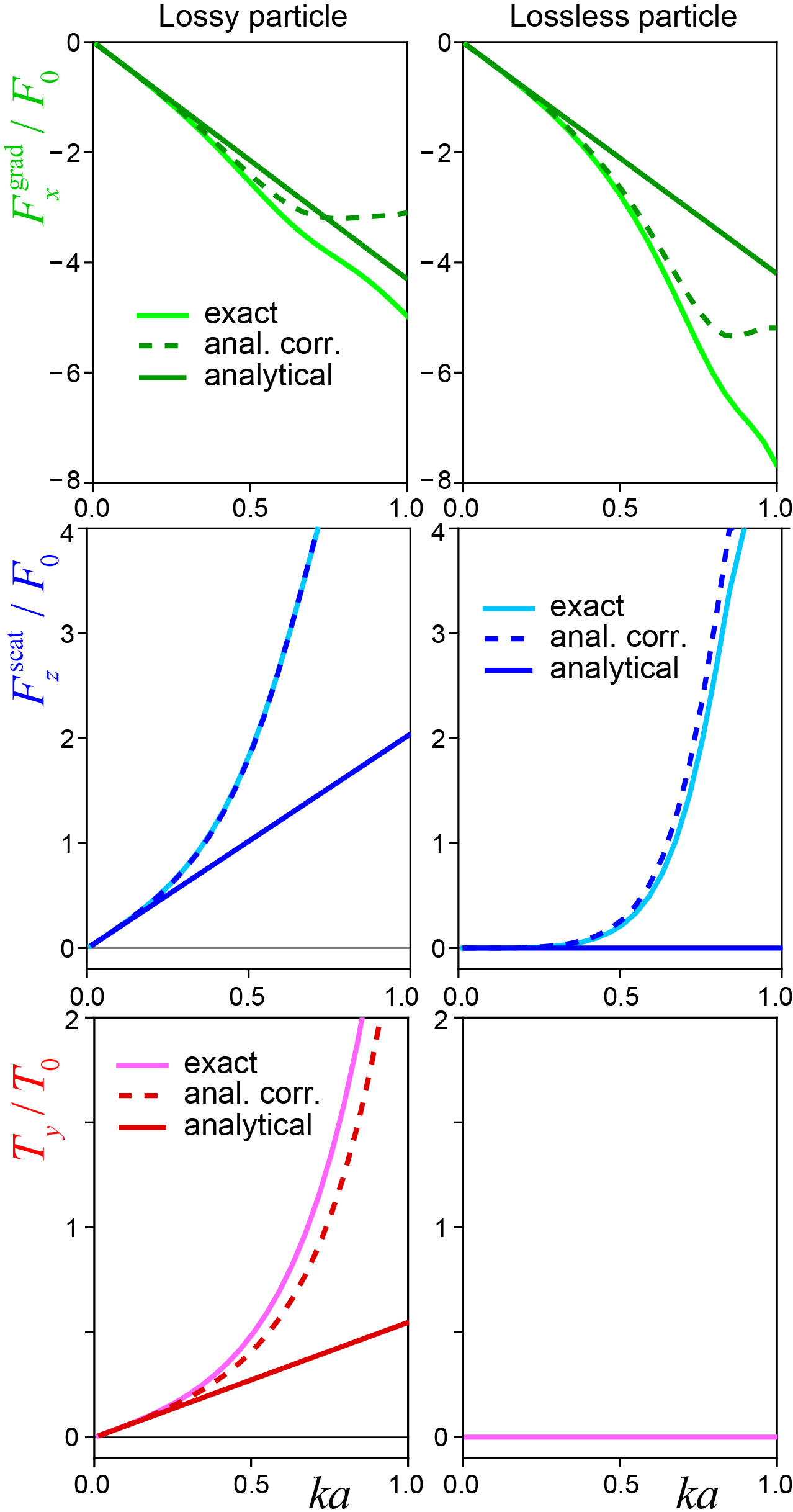}
\caption{Exact numerical and approximate analytical calculations of the gradient force, scattering (radiation-pressure) force, and torque on a spherical particle in the acoustic evanescent field, Fig.~\ref{Fig2}. The field and particle parameters are: $k_z/k = 1.34$, $\kappa/k= 0.89$, $\bar{\rho} = 2 + 0.5 i$, $\bar{\beta} = 3 + 0.7 i$ (the imaginary parts are omited in the lossless-particle case). See discussion in the text.}
\label{Fig3}
\end{figure}

Figure \ref{Fig3} shows the dependences of these two forces and torque in the field (\ref{Eq_evanescent}) on the dimensionless particle radius $ka$ for the cases of absorptive and lossles particles. We plot analytical results from Eqs.~(\ref{Eq_gradient})--(\ref{Eq_torque}), valid only to leading order, $\propto\! (ka)^3$, and the exact numerical calculations using the Mie scattering solutions together with the momentum and angular momentum fluxes, similar to the Maxwell stress tensor approach in optics (see Supplemental Material \cite{SM}). Note that the forces and torque are normalized by $F_0 = \pi \beta |A|^2 a^2 /2$ and $T_0 = F_0 / k$, so the analytical dependences (\ref{Eq_gradient})--(\ref{Eq_torque}) are linear in Fig.~\ref{Fig3}. For an absorptive particle, the analytical approximation agrees with the exact calculations for $ka \lesssim 0.3$. 

To improve the accuracy of the analytical expressions (\ref{Eq_gradient})--(\ref{Eq_torque}), one can use the exact expressions for the Mie scattering coefficients $a_0$ and $a_1$ in Eqs.~(\ref{Eq_polarizabilities}) (see Supplemental Material \cite{SM}). In this case, the monopole and dipole terms include all orders in $ka$, although the higher-order multipole terms are still neglected. The corresponding refined analytical dependences are shown in Fig.~\ref{Fig3} by dashed curves, and these agree with the exact numerical calculation for $ka \lesssim 0.8$. 

Note important peculiarities (also known in optics) of the scattering force and torque on {\it lossless} particles. First, the scattering (radiation-pressure) force vanishes only in the $(ka)^3$ order but is generally {\it non-zero} (see Fig.~\ref{Fig3}). The higher-order radiation-pressure force originates from the so-called ``radiation friction'' effect, which is described by small higher-order imaginary parts in the monopole and dipole polarizabilities \cite{Simpson2010,Albaladejo2010,Nieto-Vesperinas2010}, and also from the interference between the monopole and dipole fields \cite{Nieto-Vesperinas2010}. Using the Mie coefficients $a_0$ and $a_1$, we find that the higher-order imaginary parts of the polarizabilties can be written as $\tilde\alpha_{m} \simeq \alpha_{m} + \dfrac{ik^3}{4\pi} \alpha_{m}^2$ and $\tilde\alpha_{d} \simeq \alpha_{d} + \dfrac{ik^3}{12\pi} \alpha_{d}^2$, where $\alpha_{m,d}$ are the leading-order polarizabilities (\ref{Eq_polarizabilities}) (see Supplemental Material \cite{SM}).  
Second, the radiation torque {\it vanishes exactly} for lossless spherical particles of any radius (Fig.~\ref{Fig3}). This is also similar to optics, where the radiation-friction effect produces  the force but not the torque on the particle \cite{Nieto-Vesperinas2015,Marston1984}. Thus, the simplest analytical approximation (\ref{Eq_polarizabilities}) and (\ref{Eq_torque}) coincides with the exact calculations in this case.

{\it Conclusion.---}
We have presented a general theory of the interaction of a monochromatic acoustic wave field with a small absorbing spherical particle. Our theory is based on the complex monopole and dipole polarizabilities of the particle, and it provides simple analytical expressions for the absorption rate, radiation forces (including the gradient and scattering forces), and radiation torque. Most importantly, these expressions reveal close connections with the fundamental local properties of the acoustic field: its energy, canonical momentum, and spin angular momentum densities \cite{Shi2019,Bliokh2019,Bliokh2019a}. Thus, one can now use acoustic forces and torques to measure the canonical momentum and spin densities of sound waves, and vice versa: use canonical momentum and spin to predict radiation forces and torques. Our work also unifies theoretical approaches to the acoustic and optical field-particle interactions, and reveals close parallels between these. This provides a more fundamental  understanding and new physical insights into these important problems. 


\vspace*{0.2cm}

\begin{acknowledgments}
We are grateful to Y. P. Bliokh, A. Y. Bekshaev, and Y. S. Kivshar for fruitful discussions. This work was partially supported by MURI Center for Dynamic Magneto-Optics via the Air Force Office of Scientific Research (AFOSR) (FA9550-14-1-0040), Army Research Office (ARO) (Grant No. Grant No. W911NF-18-1-0358), Asian Office of Aerospace Research and Development (AOARD) (Grant No. FA2386-18-1-4045), Japan Science and Technology Agency (JST) (Q-LEAP program, and CREST Grant No. JPMJCR1676), Japan Society for the Promotion of Science (JSPS) (JSPS-RFBR Grant No. 17-52-50023, and JSPS-FWO Grant No. VS.059.18N), RIKEN-AIST Challenge Research Fund, the John Templeton Foundation, the Foundation for the Advancement of Theoretical Physics and Mathematics ``Basis'', and the Australian Research Council.
\end{acknowledgments}


\bibliography{bib_1}




\newpage






\begin{figure}
 \centering 
 \includegraphics[scale=0.8]{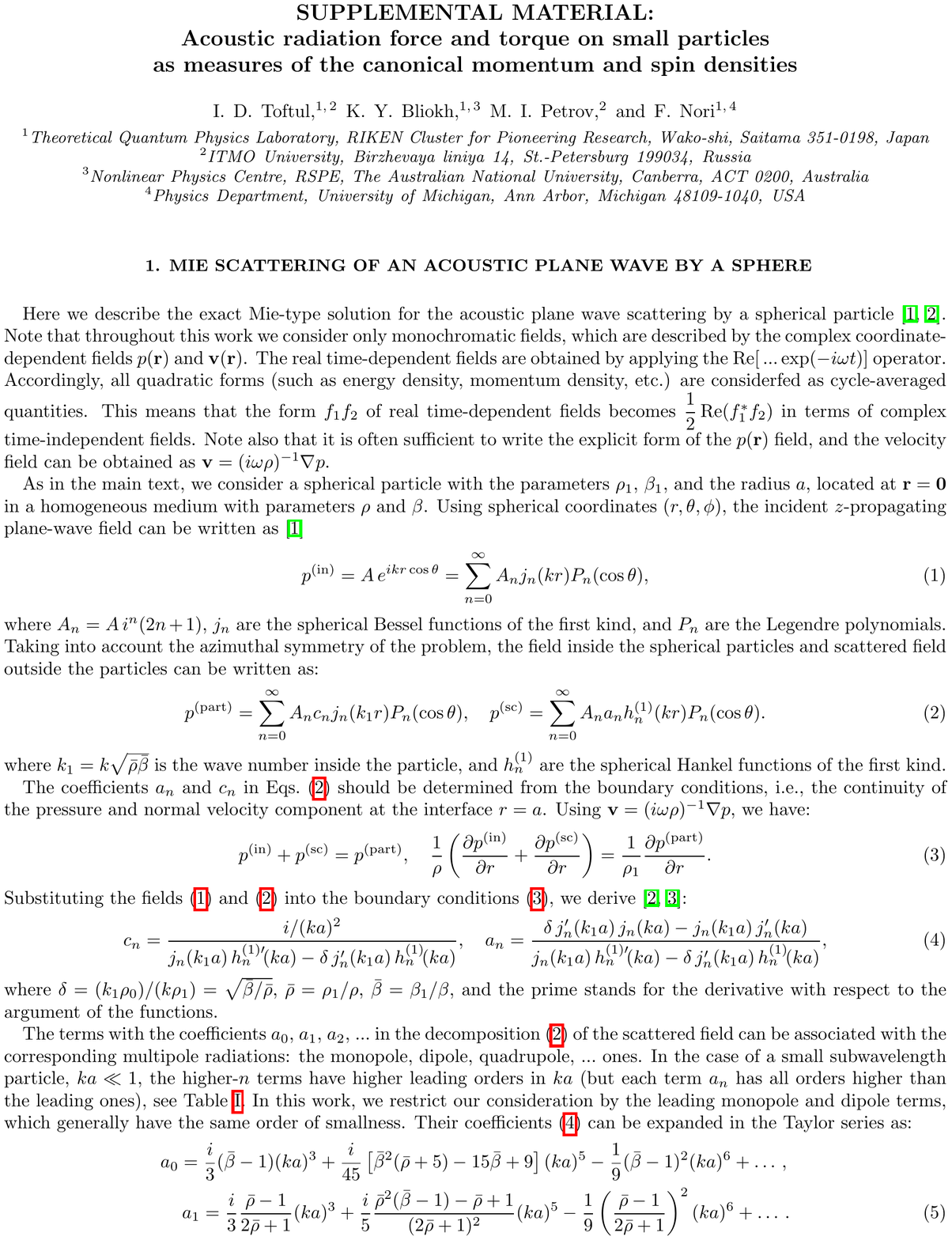}
\end{figure}

\begin{figure}
 \centering 
 \includegraphics[scale=0.8]{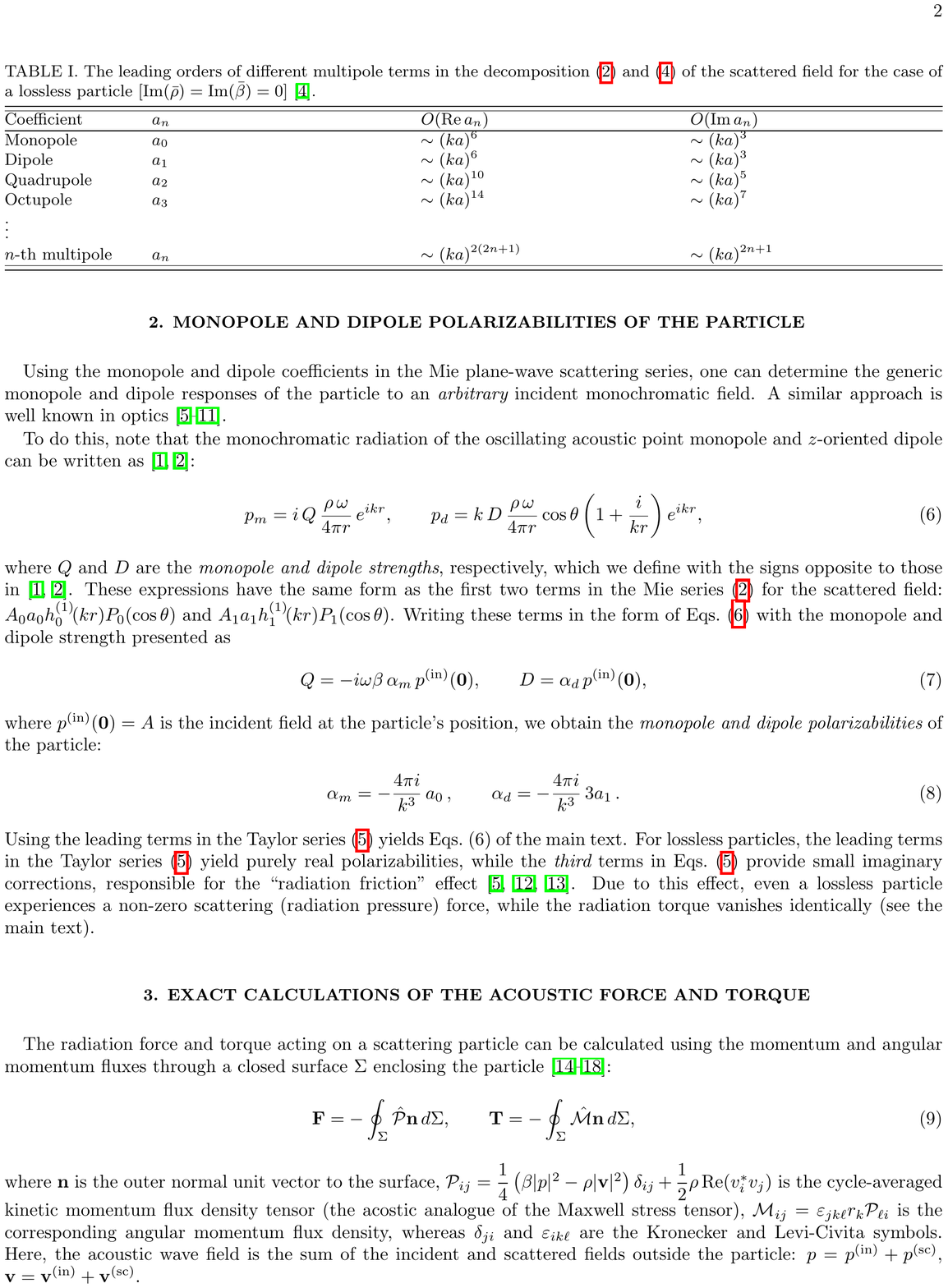}
\end{figure}

\begin{figure}
 \centering 
 \includegraphics[scale=0.8]{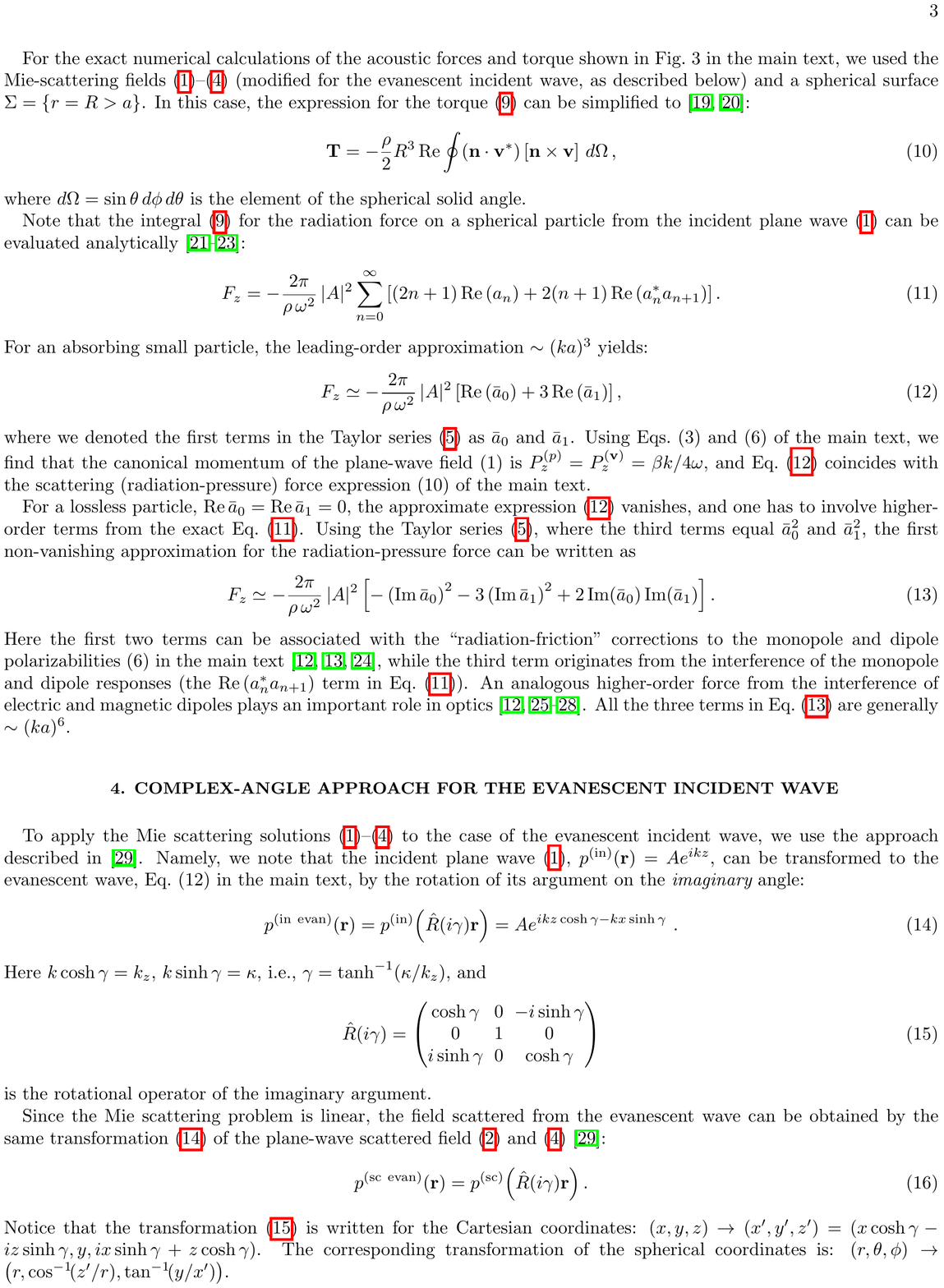}
\end{figure}

\begin{figure}
 \centering 
 \includegraphics[scale=0.8]{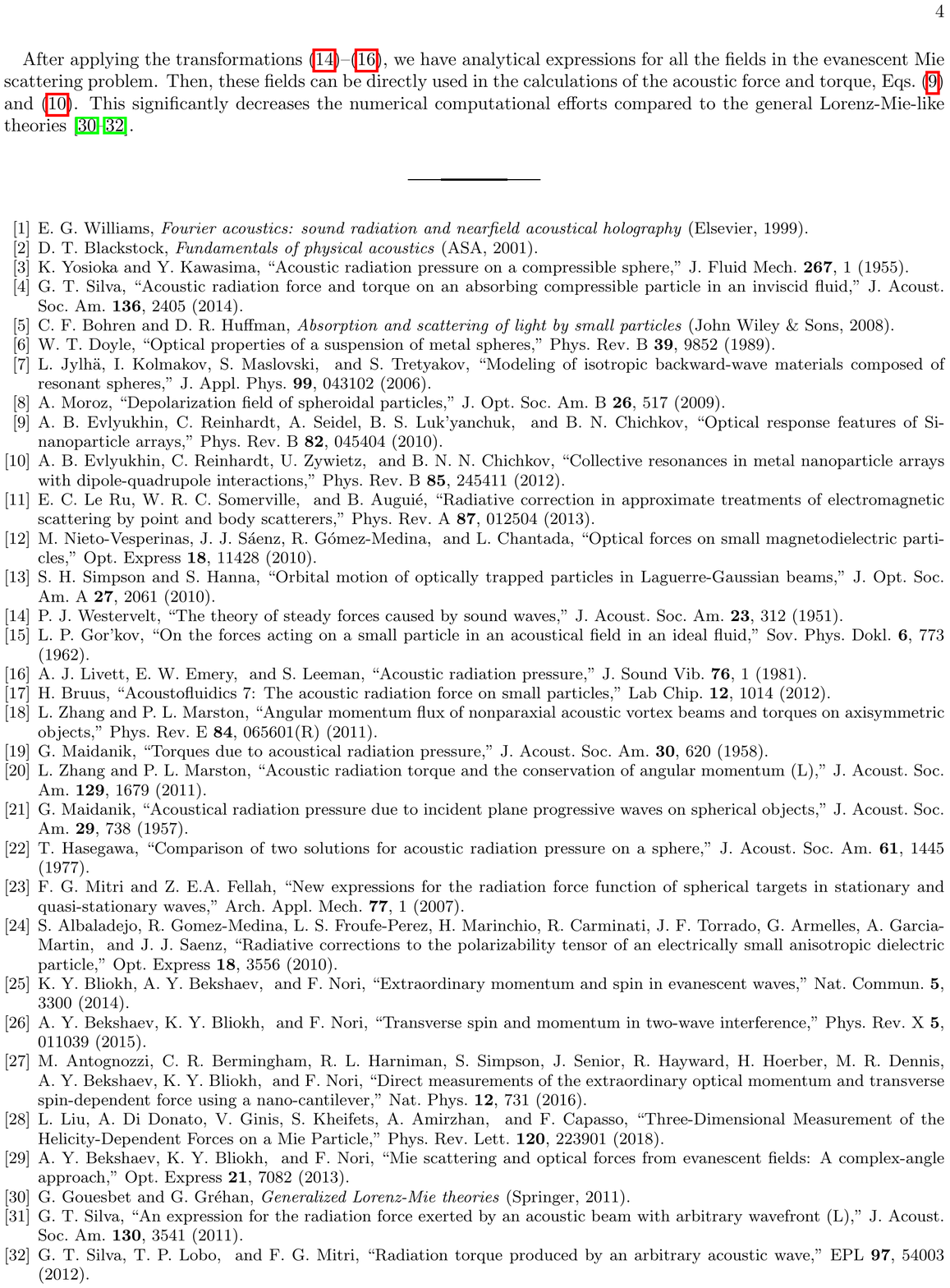}
\end{figure}

\end{document}